 \documentclass[final,3p,times,twocolumn]{elsarticle}

 \usepackage{graphics}
 \usepackage{graphicx}
 \usepackage{epsfig}

\usepackage{amssymb}
 \usepackage{amsthm}

\usepackage{lineno}
\usepackage[ruled,vlined]{algorithm2e}

\usepackage{amssymb}
\usepackage[centertags]{amsmath}
\usepackage{color}

\journal{Computer Networks}

\begin{document}

\begin{frontmatter}


\title{On-Demand Routing Algorithm with Mobility Prediction in the Mobile Ad-hoc Networks\tnoteref{t1}}
\tnotetext[t1]{This work was supported by the 2013 Research Fund of University of Ulsan}
\author{Trung~Kien~Vu}
\ead{trungkienvu@icloud.com}
\author{Sungoh~Kwon\corref{cor1}}
\ead{sungoh@ulsan.ac.kr}
\cortext[cor1]{Corresponding author}
\address{School of Electrical Engineering \\University of Ulsan \\Ulsan, Korea, 680-749}



\begin{abstract}
In this paper, we propose an ad-hoc on-demand distance vector routing algorithm for mobile ad-hoc networks taking into account node mobility. Changeable topology of such mobile ad-hoc networks provokes overhead messages in order to search available routes and maintain found routes. The overhead messages impede data delivery from sources to destination and deteriorate network performance. To overcome such a challenge, our proposed algorithm estimates link duration based neighboring node mobility and chooses the most reliable route. The proposed algorithm also applies the estimate for route maintenance to lessen the number of overhead messages. Via simulations, the proposed algorithm is verified in various mobile environments. In the low mobility environment, by reducing route maintenance messages, the proposed algorithm significantly improves network performance such as packet data rate and end-to-end delay. In the high mobility environment, the reduction of route discovery message enhances network performance since the proposed algorithm provides more reliable routes.
\end{abstract}


\end{frontmatter}


\section{Introduction}
Mobile ad hoc networks (MANETs)~\cite{4603864, citeulike:705750} consist of a set of wireless mobile nodes, which dynamically exchange data among themselves without the reliance on any fixed infrastructure. Because of the easiness to deploy and to extend networks, MANETs application scenarios include emergency and rescue operations, conference or campus settings, car networks, personal networking, and others areas. Due to limitation of transmission ranges and infrastructure-free networks, each node in such networks has responsibility not only for discovering new routes but also for relaying messages.

The most challenging problem of MANETs is how to adapt the mobility of the nodes consisting of a network. The adaptability to the mobility affects the performance of the network~~\cite{Camp02asurvey, 4141345, 1208920}. Due to changeable topology, routes from sources to destinations may be broken so nodes should efficiently discover other routes to deliver data, referred to as route discovery. After finding routes, nodes also need to efficiently maintain routes whether the discovered routes are alive or not. If a node believes that a route is alive but the route is broken and route repair was unsuccessful, when the node transfer data via the route, data will be lost and the source node initiates route discovery. Such data loss and additional overhead messages degrade network performance such as packet delivery rate and packet delay, and consume additional energy, which can be critical to a battery-operated network. Hence, routing protocol is an importance issue in MANETs for delivering the data from one node to another node. To overcome this problem there is a challenging task to develop an efficient routing protocol in MANETs.

Reactive protocols~\cite{760423} have been proposed for MANETs to adapt quickly to changing topology: the ad-hoc on-demand distance vector (AODV)~\cite{Perkins97ad-hocon-demand} routing protocol. To find a possible route, a source floods a routing request message over the network and discovers a route based on principle of the shortest path. To reduce the number of overhead messages for a route discovery, using location information, location-aided routing (LAR) protocol limits the search area~\cite{5307033}. However, such routing algorithms do not consider life times of links and routes, which are affected by node mobility, even though routing performance depends on life time of each links, referred to as link duration~\cite{1683492, 1524593}. The link duration plays an important role in the route maintenance~ \cite{1542780, 1561225, 5633523}.

In~\cite{4067771, 4346547}, the authors took into account path duration to select routes. They modeled a link duration as an exponential distribution and proposed an AODV-based routing algorithm, which chooses a route the largest expected duration. However, their statistical approach does not reflect practical link durations of individual links so the algorithm is fit to only large size networks with the random way point mobility model.

In this paper, we propose a routing algorithm taking into account link durations and life times of routes, and a maintenance scheme to reduce the number of overhead to maintain routes. To that end, each node predicts link durations based on measured locations of neighboring nodes. Based on the predicted link durations, nodes choose the most reliable route and adapt periods for route maintenance so as to reduce the amount of redundant overhead message in the network. Moreover, nodes adapt time duration for maintaining routes according to mobility.



The rest of this paper is organized as follows. In Section~\ref{Sys-Model}, we describe the system model and problems to study. In Section~\ref{Proposed-algorithm}, we proposed a routing algorithm. In Section~\ref{Simulation}, we provide numerical results to study the efficacy of the scheme and conclude the paper in Section~\ref{Conclusion}.
\section{System Model and Problem}
\label{Sys-Model}
\subsection{System model}
\label{Assumptions}
In this paper, we consider an ad-hoc network consisting of $n$ mobile nodes with transmission range~$r$. Each node~$i$ is assumed to employ a location measurement device such as the global positioning system~(GPS) \cite{6295661} and measure its own position $X_{i}(t)$ and velocity $\overrightarrow{V_{i}}(t)$ at any time~$t$.

The distance $D_{(i,j)}(t)$ between nodes~$i$ and $j$ at time~$t$ is defined as $D_{(i,j)}(t)$ = $|X_{i}(t)-X_{j}(t)|$,
where $|X|$ stands for a Euclidian distance of vector $X$. A link between nodes~$i$ and $j$ is denoted as $(i,j)$.
Link$~(i,j)$ is called \emph{valid} or \emph{connected} link at time~$t$ when the distance between nodes~$i$ and $j$ is
less than the transmission range~$r$, i.e., $D_{(i,j)}(t) \leq r$. Otherwise, \emph{broken} or \emph{disconnected}. The
link duration of link~$(i,j)$ is defined as the time interval while the link is valid.

Due to limited transmission range, packets are delivered from a source to a destination in a multi-hop manner via a route,
which is defined as a set of links. For given source and destination nodes, $s$ and $d$,
$K$ possible routes at time $t$ are denoted as $R_{(s,d)}^{(k)}(t)$ for $k \in \{1, \cdots, K\}$, which consist of $\left|
R_{(s,d)}^{(k)}(t) \right|$ links.
\subsection{Summary of AODV Routing Protocol}
\label{Routing-Protocol}
The AODV routing algorithm is a reactive routing protocol, which is designed for mobile ad-hoc networks to adapt the movement of mobile node as well as a variety of data traffic levels. There are two importance mechanisms: route discovery and route maintenance.

The route discovery is initiated by a source node that has data to send a destination node or the source node does not have an active route in its routing table. To find a route to the destination, the source node broadcasts a route request message~(RREQ) including a sequence number to neighboring nodes. The RREQ message is flooded through the entire network until the message reaches the destination or an intermediate node. Each node receiving the RREQ message stores a reverse route to the source and broadcasts the message to their neighboring nodes if the node is not the destination and the message is not a duplicate. When the RREQ message arrives at a destination node or intermediate node that has a valid route to the destination, the node sends a route reply message~(RREP) to the neighboring node in a reverse route in a unicast manner. The RREP message contains the number of hops to reach the destination node and the sequence number for the destination. A node receiving the RREP message forward to the source backward in the stored reverse route and creates or updates a forward route to the destination.

In order to maintain routes, nodes periodically send a hello message to their neighbors to check if links are connected. If a
node does not receive any hello message from its neighbors during a certain time
period, referred to as life time of hello message, the node assumes that the link to the neighbor is currently disconnected and
reports the link failure to the source corresponding to the link via a route error~(RRER) message. AODV also uses the link local detection method to repair link breakages.


\subsection{Problem}
\begin{figure}
  \centering
  \includegraphics[width=3.5in]{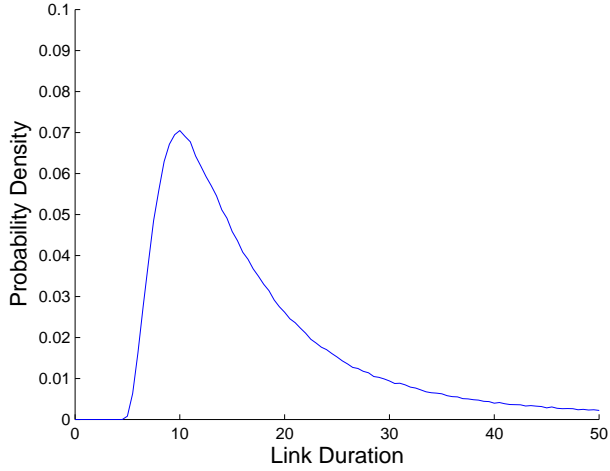}\\
  \caption{Probability density of link duration}\label{ld}
\end{figure}

The original AODV selects a route with fixed lifetime instead of any reliable lifetime parameters and uses the route until a link composed of the route is broken. The original AODV also uses a fixed value as a duration for route maintenance. Such fixed settings may not be appropriate to MANET, since each link has different link duration as well as active route duration due to node mobility, as shown in Fig.~\ref{ld}, which shows the probability density of link duration when nodes with radius 300~m move with randomly chosen speeds between 4~m/s and 24~m/s and randomly chosen directions between 0 and 2$\pi$.

The discrepancy between real link durations and lifetime parameters induces more overhead messages and more delivery failure, which results in performance degradation. For example, in the case when nodes move slowly, active link duration can be longer than the prefixed time-out so that the route discovery due to time expiration is redundant and increase overhead message over the network. In the case of high mobility of nodes, the pre-fixed timeout can be shorter than the actual link duration so that packet delivery fails due to route failure. The delivery failure initiates the route recovery routine to find a route from the source to the destination, as described in Section~\ref{Routing-Protocol}. The process to reconstruct a new route results in additional maintenance messages over the network and increases end-to-end packet delays.

By estimating a link duration of each link, an effective routing algorithm may be necessary for MANETs and we will propose a mobility prediction-based AODV routing protocol.

\section{Proposed Algorithm}
\label{Proposed-algorithm}
In this section we propose a routing protocol with adaptive time-out parameters. To that end, we introduce a link duration estimate, route expiration time, which are followed by a proposed routing algorithm\footnote{Other published results of our work have been published in~\cite{vu2014mobility, vu2014impact}}.
\subsection{Estimated link duration and route expiration time}
For given measured location~$X_{i}(t)$ and velocity~$\overrightarrow{V_{i}}(t)$ of node~$i$ at current time~$t$, the
node can estimated its future location~$\tilde{X}_{i}(t+\Delta t)$ at time~$t+\Delta t$, defined as equation (\ref{eq:solve})
that is presented in \cite{1589667}.
\begin{equation}\label{eq:solve}
  \tilde{X}_{i}(t+\Delta t) = X_{i}(t) + \overrightarrow{V_{i}}(t) \Delta t
\end{equation}

\begin{figure}
  \centering
  \includegraphics[width=3.5in]{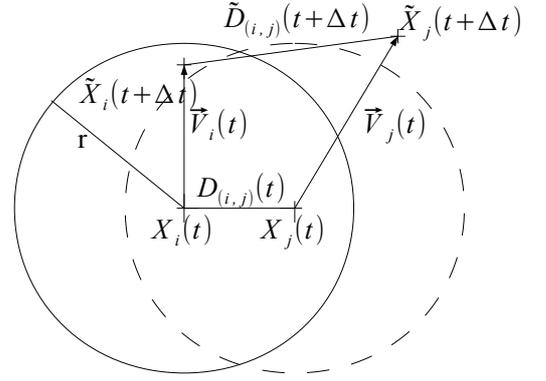}\\
  \caption{Estimated link duration}\label{s1}
\end{figure}

In the case of MANET, all the nodes are mobile and the connectivity of two nodes depends on the \emph{relative mobility} of the two nodes such as the relative location~$X_{(i,j)}(t)$ and relative velocity~$V_{(i,j)}(t)$ of node~$i$ with respect to
node~$j$ at current time~$t$, which are defined as $X_{(i,j)}(t) = X_{i}(t)-X_{j}(t)$ and $V_{(i,j)}(t) = \left(\overrightarrow{V_{i}}(t)-\overrightarrow{V_{j}}(t)\right)$, respectively. Hence, as displayed in Fig.~\ref{s1}, the estimated distance between nodes~$i$ and $j$ elapsed time~$\Delta t$ from current time~$t$ can be expressed as
\begin{eqnarray*}
  \tilde{D}_{(i,j)}(t+\Delta t) &=& \left| \tilde{X}_{i}(t+\Delta t) - \tilde{X}_{j}(t+\Delta t) \right| \\
  &=& \left| X_{(i,j)}(t) + \overrightarrow{V_{(i,j)}}(t) \Delta t \right|
\end{eqnarray*}

Only if the distance of two nodes are less than or equal their transmission range, the link between the two nodes is connected,
as defined in Subsection~\ref{Assumptions}. Hence, the estimated link duration
while the link between nodes~$i$ and $j$ is connected is denoted as $LDT_{(i,j)}$ and defined as
\begin{eqnarray}
  &LDT_{(i,j)} = & \min \Delta t \nonumber \\
  &\mathrm{subject \ to} & \tilde{D}_{(i,j)}(t+\Delta t) \leq r, \label{eq-ldt}
\end{eqnarray}
where $r$ is the transmission range.

For given route $R_{(s,d)}^{(k)}$ between nodes~$s$ and $d$, we define the route expiration time~$RET_{(s,d)}^{(k)}$ as the
least link duration time of the route, which is expressed as
\begin{eqnarray}
  RET_{(s,d)}^{(k)} = \min_{(i,j) \in RET_{(s,d)}^{(k)}} LDT_{(i,j)}. \label{eq-ret}
\end{eqnarray}

The link duration times of links are locally measured at each nodes and the route expiration time of each route is delivered to
the destination via {\it route expiration time} parameter in the AODV routing
algorithm.
\subsection{Proposed Algorithm for Route Discovery}
\label{Algorithm-RD}
When new data arrive at a node, the source node finds an active route to the corresponding destination in its routing table, as in Subsection~\ref{Routing-Protocol}. If no active route exists, the source node initiates route discovery to find a route to the destination node by broadcasting a RREQ message to neighboring nodes. The RET field and the hop count field in the RREQ message are initially set to the infinity and one, respectively.

\begin{figure}
  \centering
  \includegraphics[width=3.5in]{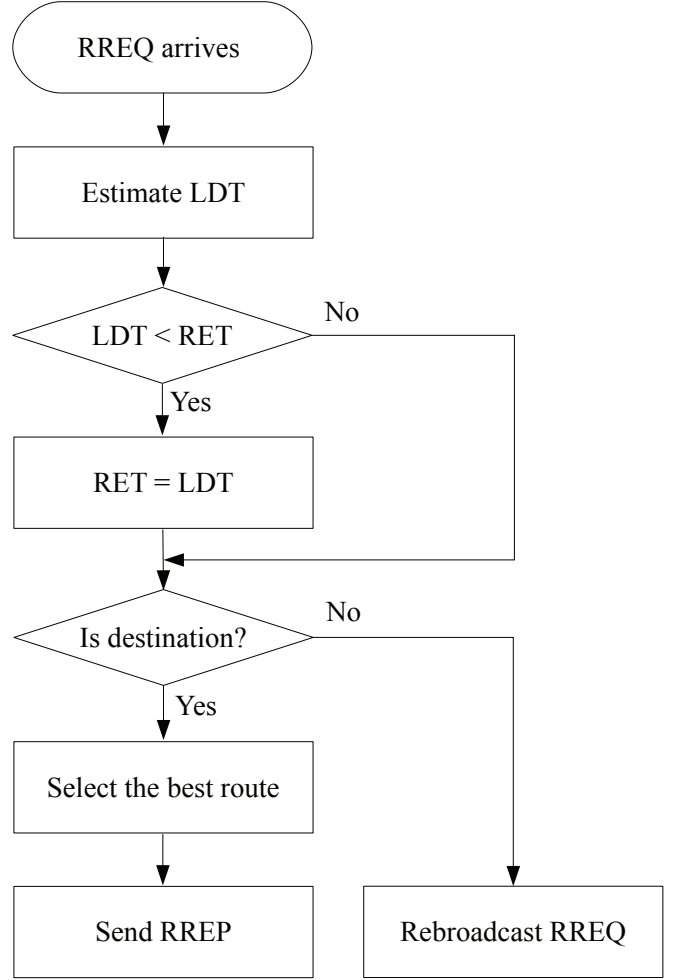}
  \caption{Actions of an intermediate node.}\label{action1}
\end{figure}

On reception of RREQ, nodes compute the link duration time~(LDT) value between the RREQ sender and itself, which implies the estimated life time of the link, from~(\ref{eq-ldt}). The LDT value will be compared with a RET value in the RREQ message. If the LDT is smaller than the RET in RREQ, then the receiving nodes update the RET value with the new LDT and check if they are the destination of the RREQ comparing their destination addresses and its own address. If it is not the destination, the nodes broadcast the receiving RREQ to other nodes after increasing the hop count by one. The actions of nodes receiving RREQ are showed in Fig.~\ref{action1}.

In the case when a node is the destination of RREQ, then the node waits for time interval~$T_{w}$ and collects RREQs whose destination is the node. After the time interval~$T_{w}$, the destination decides the best route among the received routes and replies a RREP message setting the lifetime field as the corresponding RET. Nodes receiving RREP relay the message in a unicast way until the message reach at the source, as described in Subsection~\ref{Routing-Protocol}.

To decide the best route among candidate routes, we can consider three different ways based on the number of hops and the
amount of route expiration time as follows.
\subsubsection{Longest RET among the shortest path}
\label{LRETSP}
The destination node collects RREQs for time duration~$T_{w}$ after receiving the first RREQ. Among the received RREQs, the destination node choose the route of which hops is the smallest. If there exist multiple routes whose hops is the same as the smallest number of hops, then the destination selects the route whose RET is the greatest among the candidates.
\subsubsection{Longest RET among the candidates}
\label{LRETAC}
The second option is to simply choose the longest RET as a route among the received RREQs for time duration~$T_{w}$ since a route having the longer RET is more stable and longer to live. The selection can be expressed as
\[ \mathrm{argmax}_{k \in K_{T_{w}}} RET_{(s,d)}^{(k)},\]
where $K_{T_{w}}$ is a set of collected routes during~$T_{w}$.
\subsubsection{Largest ratio of RET to hops}
\label{LRRET2H}
The algorithm in Section~\ref{LRETSP} is restricted by the smallest hops so that the longest RET route cannot be chosen. The
algorithm in Section~\ref{LRETAC} considers only the RET so that the route of too many
hops can be selected. Hence, the other option is to choose the route of the largest ratio of the RET to the number of hops,
which is expressed as
\[ \mathrm{argmax}_{k \in K_{T_{w}}} \frac{RET_{(s,d)}^{(k)}}{\mathrm{the \ number \ of \ hops}}.\]
After selecting the route, the destination sends an RREP message to the source in a unicast manner and the reverse way.

\subsection{Proposed Algorithm for Route Maintenance}
\label{Algorithm-RM}
In order to maintain routes, nodes periodically send a hello message to their neighbors to check the connectivity. In the proposed algorithm, the frequency is also adaptively chosen according to the link duration time instead of the fixed period~(1 second) since the redundant hello messages increase the routing overhead. The frequency for route maintenance between nodes~$i$ and $j$ is defined as
\[\max\left\{1,\frac{\min LDT_{(i,j)}}{\alpha} \right\},\]
where $\alpha$ is a control parameter.

\subsection{Example}
\begin{figure}
  \centering
  \includegraphics[width=3.5in]{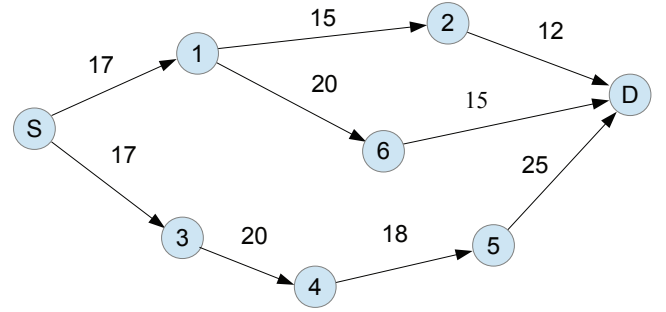}
  \caption{An Example of LDT and RET}\label{s2}
\end{figure}
In this subsection, we describe the proposed algorithm with an example. There exist eight nodes including source and destination nodes, as in Fig.~\ref{s2}. When node~$s$ has data to deliver to node~$d$ and there is no active route in the routing table, node~$s$ initiates a route discovery process. After setting the number of hop, RET, and the destination node as zero, infinity, and node~$d$, respectively, Node~$s$ broadcasts a RREQ message to its neighbors.

Nodes~$1$ and $3$ that receive a RREQ message measure $LDT_{(s,1)}$ and $LDT_{(s,3)}$ before updating $RET_{(s,d)}$ in
(\ref{eq-ret}) at each node. If nodes are not the destination node in RREQ, the nodes
broadcast the RREQ message to its neighborhood after updating corresponding parameters.

If a node is the destination node in RREQ, the destination node~$d$ waits for time~$T_{w}$ and gathers RREQs via other routes. Based on the gathered RREQs, the destination choose a route and send RREP to
the source in the reverse order of the selected RREQ.

The original AODV will select Route~1 that is the route in the first arrived RREQ. However, the proposed algorithms in
Section~\ref{LRETSP} and Section~\ref{LRRET2H} will choose Route~2 while Route~3 will be selected by the
proposed algorithms in Section~\ref{LRETAC}. Table ~\ref{tab-1} summarizes the example.
\begin{table}
\caption{Route expiration time} 
\centering 
\begin{tabular}{c c c c} 
\hline 
Route &     & Hop count & RET(second)\\
\hline 
 Route 1& $s \rightarrow 1 \rightarrow 2 \rightarrow d$                & 3              & 12 \\ 
 Route 2& $s \rightarrow 1 \rightarrow 6 \rightarrow d$                & 3              & 15 \\ 
 Route 3& $s \rightarrow 3 \rightarrow 4 \rightarrow 5 \rightarrow d$  & 4              & 17 \\ 
\hline 
\end{tabular}
\label{tab-1} 
\end{table}
\section{Performance Evaluation}
\label{Simulation}
The routing protocol is implemented in NS-2 network simulator, which is a discrete event simulator developed by University of California at Berkeley and the virtual inter-network testbed (VINT) project \cite{Varadhan03thens}. For simulations, there are 100 nodes initially distributed in an area of 2~km by 1.5~km with 250~m of transmission range. We run simulations with twenty difference random seeds and average the simulation results.

The Random Waypoint Mobility (RWP) \cite{hyytia2005random} model is used as a referenced mobility model, in which mobile nodes move from their current locations to new locations by randomly choosing directions and speeds. Upon arrival at a destination, after a pause time, they choose another random destination in the simulation area and travel toward the destination with an uniformly distributed speed between the maximum speed and minimum speed. We set the pause time to zero to represent constant mobility.

The constance bit rate (CBR) traffic under the user datagram protocol (UDP) is used to compare accurately different routing protocols with sending rate of 5 packets per second and 512 byte of packet size. The parameter settings are listed in Table ~\ref{parameter}.
\begin{table}
\caption{Parameter settings} 
\centering 
\begin{tabular}{c c} 
\hline 
Parameter & Values \\ 
\hline 
Network simulator                   & NS-2.34 \\ 
Simulation area                     & 2 km $\times$ 1.5 km \\ %
Number of mobile nodes              & 100 \\
Simulation time                     & 900 s \\
Mobility model                      & Random way point \\
Pause time                          & 0 s\\
Packet generation rate              & 5 packets/second \\
Packet size                         & 512 byte \\
Transmission range                  & 250 m \\ 
\hline 
\end{tabular}
\label{parameter} 
\end{table}

In simulations, we compare our proposed routing algorithm with the original AODV. For the comparison, three different route selection schemes~in Section~\ref{Algorithm-RD} are considered and named as AMP-AODV I, AMP-AODV II, and AMP-AODV III.

Three metrics are considered to evaluate the network performance according to \cite{Broch98aperformance, 832168}: the amount of overhead packets, the packet delivery ratio, and the end-to-end delay. For the amount of overhead packets, we count the number of packets for route discovery and route maintenance. For comparison, the total number of overhead packets is normalized by the number of packets successfully delivered to destinations. The packet delivery radio is defined as a ratio of the number of generated packets to the number of packets received at corresponding destinations. The average end to end is defined as the average time for packets to delivered from sources to destinations.

\subsection{Comparison of Overheads}
\label{reducing}
Ten source-destination pairs generates 5 packets per second during the simulation time and the amount of overhead packets are compared at different mobility environments:low mobility, middle mobility and high mobility. For the low mobile environment, we set nodes' speed for RWP to 1~m/s, which is a pedestrian speed~(3.6~Km/h). We also set 10~m/s and 20~m/s as the node speeds for the middle mobility and the high mobility, respectively. The overhead messages can be categorized into two: route discovery message and route maintenance message. Route request messages and route reply messages are for route discovery and hello messages and route error messages are for route maintenance.

\begin{figure}
  \centering
  \includegraphics[width=3.6in]{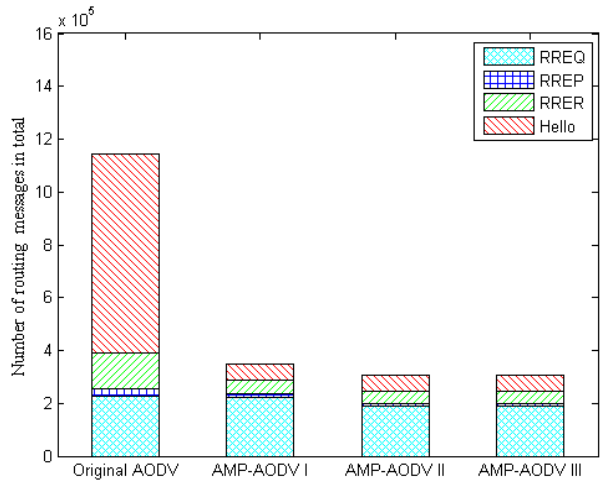}
  \caption{Route discovery and route maintenance in case of the constant 1 m/s speed}\label{sc03}
  \end{figure}

\begin{figure}
  \centering
  \includegraphics[width=3.6in]{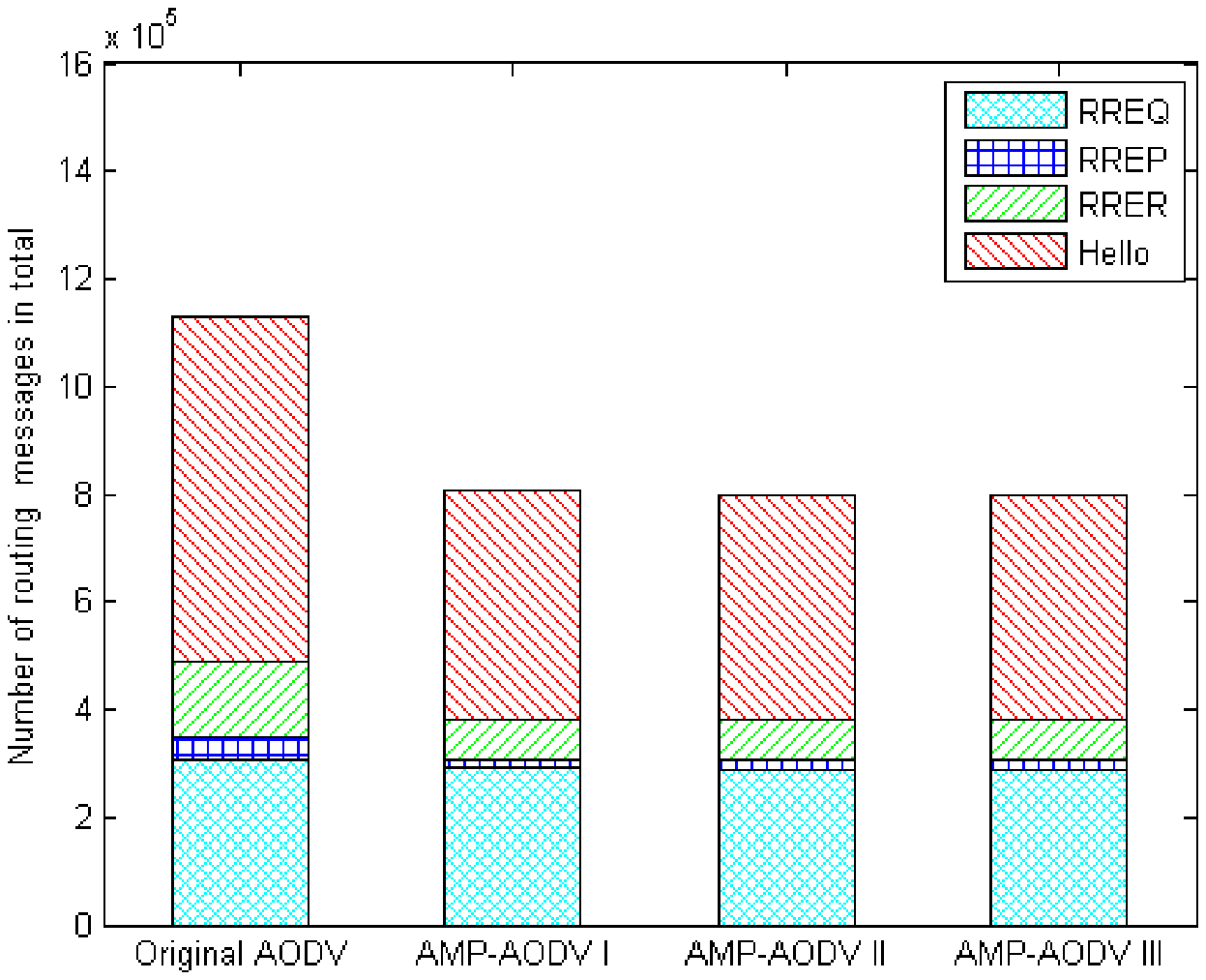}
  \caption{Route discovery and route maintenance in case of the constant 10 m/s speed}\label{sc04}
  \end{figure}

\begin{figure}
  \centering
  \includegraphics[width=3.6in]{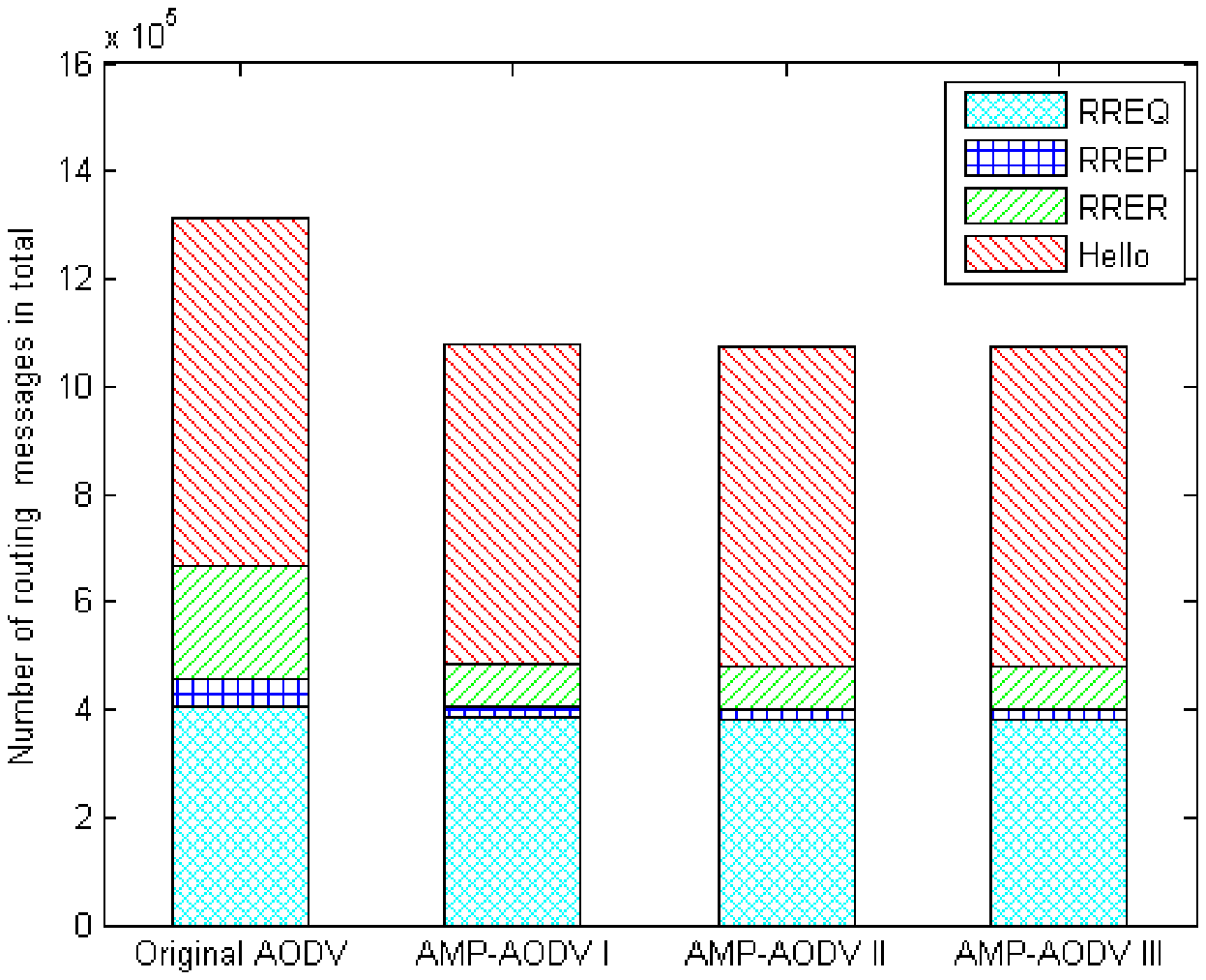}
  \caption{Route discovery and route maintenance in case of the constant 20 m/s speed}\label{sc05}
  \end{figure}

In the case of the original AODV, the amount of hello messages for maintaining routes takes a large share among overhead messages in all the scenarios, as in Figs.~\ref{sc03}, \ref{sc04}, and \ref{sc05}. Especially, the lower mobility has the more portion of hello messages compared to those of other messages. The redundant messages induce performance degradation. To reduce such redundant message, our proposed algorithm adopts an adaptive period for route maintenance and reduces a significant amount of redundant messages for maintenance in the low mobile environment.

The higher mobility induces the higher probability that active routes in nodes are actually broken due to broken links. The more frequently broken links results in the more overhead messages. As compared the original AODV in Figs.~\ref{sc03}, \ref{sc04}, and \ref{sc05}, the overhead messages for route discovery has a more portion out of the total overhead messages. Compared to the simulation results of AODV, our proposed algorithm generates much less overhead messages for route discovery.

\begin{figure}
  \centering
  \includegraphics[width=3.6in]{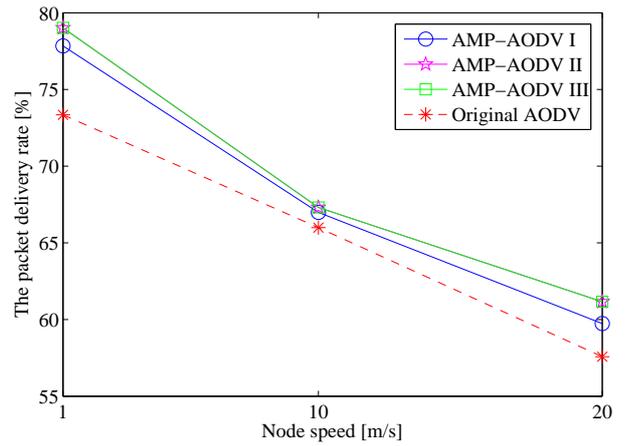}
  \caption{The packet delivery ratio versus the constant node velocity}\label{sc00}
  \end{figure}

\begin{figure}
  \centering
  \includegraphics[width=3.6in]{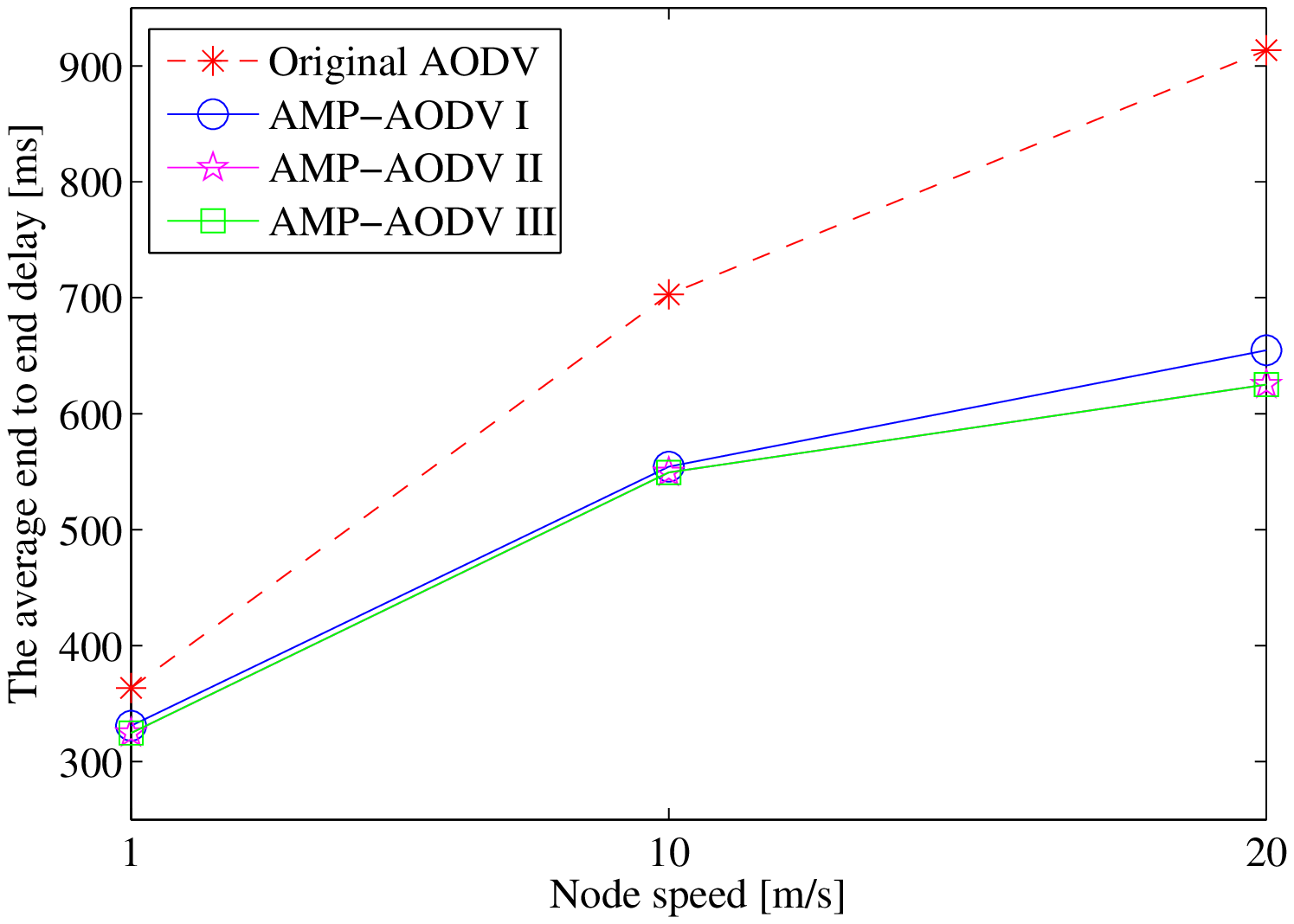}
  \caption{The average end to end delay versus the constant node velocity}\label{sc02}
  \end{figure}

Due to such redundant message reduction, the routing performance can be improved, as shown in Figs.~\ref{sc00} and \ref{sc02}. Our proposed algorithms successfully deliver more packets to destinations than the original AODV. Moreover the delivered packets of our algorithms spend less times than the original AODV. Among our algorithms, AMP-AODV~I generates more route request messages than the others, which results in less performance than the others.

\subsection{The impact of offered traffic load on performance}
In this subsection, we evaluate the impact of traffic load on the performance of routing protocols. To do that we increase the number of source-destination pairs. Each source generates CBR data with 20~Kbps and delivers the data to the corresponding destination via routes. For mobility, each node follows the RWP mobility model with randomly selected speed between 1~m/s and 20~m/s.

\begin{figure}
  \centering
  \includegraphics[width=3.6in]{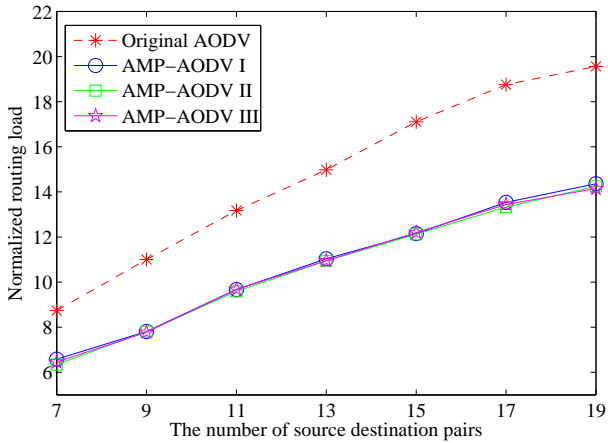}
  \caption{Normalized routing load versus traffic load.}\label{l25}
\end{figure}

\begin{figure}
  \centering
  \includegraphics[width=3.6in]{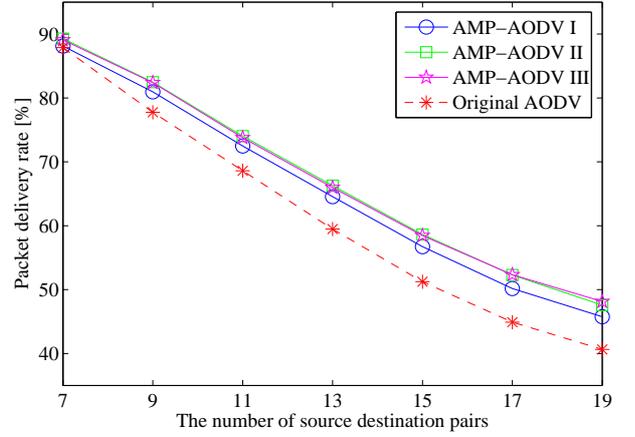}
  \caption{The packet delivery rate versus traffic load.}\label{l21}
\end{figure}

\begin{figure}
  \centering
  \includegraphics[width=3.6in]{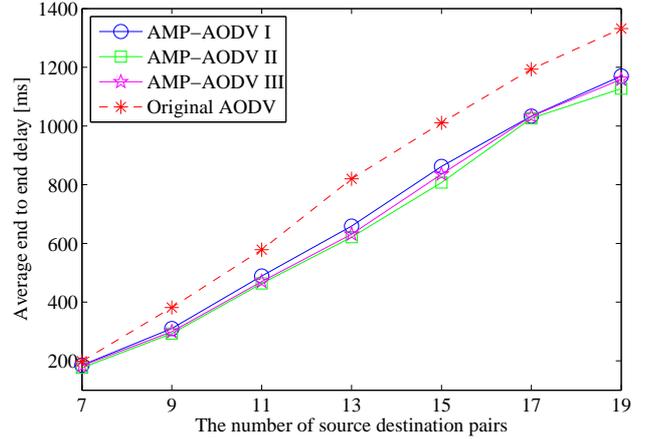}
  \caption{The average end to end delay versus traffic load.}\label{l23}
\end{figure}

Figure~\ref{l25} shows that the number of overhead messages normalized by the number of successfully delivered packets. As studied in the previous subsection, our proposed algorithms reduce the overhead messages, so that, compared to the original AODV, our proposed algorithms demand smaller number of overhead messages per packet. Such reduction of redundant messages brings out performance improvement, as shown in Figs.~\ref{l21} and \ref{l23}.

Figure~\ref{l21} shows that the packet delivery rates decrease as the number of pairs for packets increase. The more traffic load in the network induces the more packet delay due to transmission collision and congestion, as in Fig.~\ref{l23}. The more delay results in the more probability that existing links are broken due to mobility. Compared to the original AODV algorithm, our proposed algorithms estimate link life time to adapt maintenance periods and route life time to select more reliable routes so that our algorithms outperform the routing algorithm without link life time estimation.

\subsection{The impact of node density}
In this subsection, we study the impact of node density on routing performance. The number of nodes in the given area varies from 75 nodes to 175 nodes when 11 source-destination pairs are chosen and each source generate 20~kbps data for its corresponding destination.

\begin{figure}
  \centering
  \includegraphics[width=3.6in]{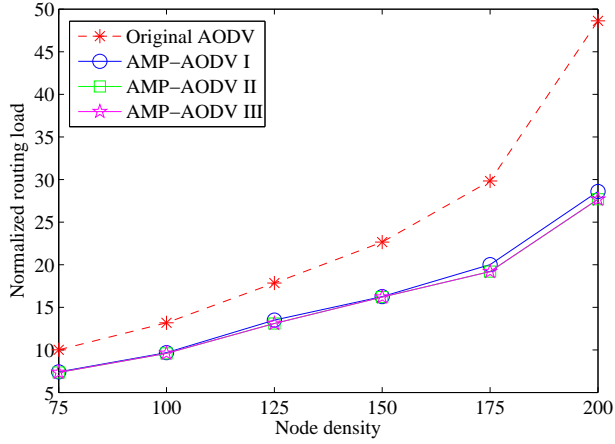}
  \caption{Normalized routing load versus node density.}\label{l45}
  \end{figure}

Figure~\ref{l45} shows that the number of overhead messages also increases as the number of nodes increases in the given area. The overhead increment is caused by the increase of the number of links to maintain in the network. Due to adaptive maintenance period, our algorithms need smaller number of overheads per packets than the original AODV.

\begin{figure}
  \centering
  \includegraphics[width=3.6in]{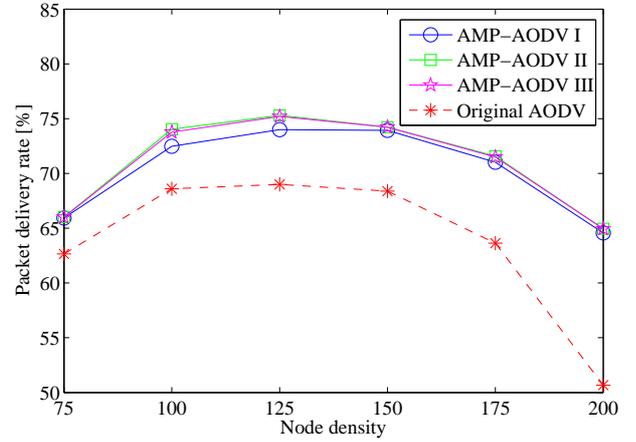}
  \caption{The packet delivery rate versus node density.}\label{l41}
\end{figure}

\begin{figure}
  \centering
  \includegraphics[width=3.6in]{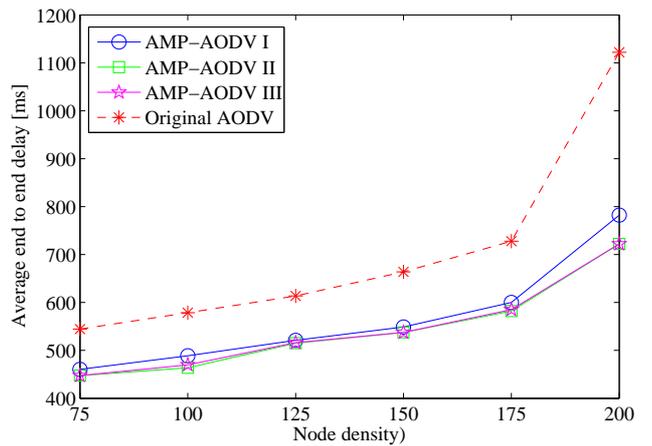}
  \caption{The average end to end delay versus node density.}\label{l43}
\end{figure}

If the number of nodes is too small, feasible routes between sources and destinations do not exist in the network. In such environments, as the number of nodes increase, the packet delivery rate increases, as in Fig.~\ref{l41}. However, behind a certain number of nodes, the larger number of nodes hinders packet delivery due to the more overhead message to maintain links, as shown in Fig.~\ref{l45}. Such delivery interference results in end-to-end delays so that our proposed algorithms improve delivery performance compared to the original AODV, as shown in Fig.~\ref{l43}.

\section{Conclusion}
\label{Conclusion}
In this paper, considering node mobility, an adaptive mobility prediction routing algorithm is proposed. The algorithm estimates link duration based on the measured mobility and chooses the best route with taking into account the link durations. Moreover the algorithm also applies the mobility estimates for route maintenance. Compared to a routing algorithm without mobility prediction, the proposed algorithm can significantly reduce the number of overhead messages for route discovery and route maintenance. For example, the proposed algorithm reduces the number of overhead messages by about one third compared to the original AODV when nodes keep moving at 1~m/s. Such reduction of overhead messages results in performance improvement such as packet delivery rate and end-to-end delay.





\bibliographystyle{elsarticle-num}
\bibliography{mybibdatabase}
\newpage






\end{document}